# Chapter Number

# Tailoring of the local environment of active ions in rare-earth- and transition-metal-doped optical fibres, and potential applications

Bernard Dussardier[1,*], Wilfried Blanc[1] and Pavel Peterka[2]
[1]*Laboratoire de Physique de la Matière Condensée, Université de Nice Sophia-Antipolis - CNRS, UMR 6622, Parc Valrose - F-06108 Nice France*
[2] *Institute of Photonics and Electronics AS CR, v.v.i., 182 51 Prague Czech Republic*

## 1. Introduction

During the last two decades, the development of optical fibre-based sophisticated devices have benefited from the development of very performant optical fibre components. In particular, rare-earth (RE)-doped optical fibres have allowed the extremely fast development of fibre amplifiers for optical telecommunications (Desurvire, 1994, 2002), lasers (Digonnet, 2001) and temperature sensors (Grattan & Sun, 2000). The most frequently used RE ions are $Nd^{3+}$, $Yb^{3+}$, $Er^{3+}$, $Tm^{3+}$ for their optical transitions in the near infrared (NIR) around 1, 1.5 and 2 µm. A great variety of RE-doped fibres design have been proposed for specific applications: depending of the RE concentration and nature of the fibre glass, various schemes (in terms of electronic transition within the RE populations) have been implemented. For example, $Er^{3+}$-doped fibre amplifiers (EDFA) for long haul telecommunications use the very efficient 1.55 µm optical transition, whereas high concentrations of $Yb^{3+}$ and $Er^{3+}$ codoped in the same fibre allowed efficient non-radiative energy transfers from the 'sensitizer' ($Yb^{3+}$) to the 'acceptor' ($Er^{3+}$) in order to increase the power yield of the system, applied in power amplifiers and lasers. All the developed applications of amplifying optical fibres are the result of time consuming and careful optimization of the material properties, particularly in terms of dopant incorporation in the glass matrix, transparency and quantum efficiency.

RE-doped fibres are made of a choice of glasses: silica is the most widely used, sometimes as the result of some compromises. Alternative glasses, including low maximum phonon energy (MPE) ones, are also used because they provide better quantum efficiency or emission bandwidth to some optical transitions of particular RE ions. The icon example is the $Tm^{3+}$-doped fibre amplifier (TDFA) for telecommunications in the S-band (1.48-1.53 µm) (Komukai et al., 1995), for which low MPE glasses have been developed: oxides (Minelly & Ellison, 2002; Lin et al., 2007), fluorides (Durteste et al., 1991), chalcogenides (Hewak et al.,

1993), etc. Although some of these glasses have a better transparency than silica in the infrared spectral region (above 2.3 µm), for applications in the near infrared (NIR) these glasses have some drawbacks not acceptable at a commercial point of view: high fabrication cost, low reliability, difficult connection to silica components and, in the case of fibre lasers, low optical damage threshold and resistance to heat. To our knowledge, silica glass is the only material able to meet most of applications requirements, and therefore the choice of vitreous silica for the active fibre material is of critical importance. However a pure silica TDFA would suffer strong non-radiative de-excitation (NRD) caused by multiphonon coupling from $Tm^{3+}$ ions to the matrix. Successful insulation of $Tm^{3+}$-ions from matrix vibrations by appropriate ion-site 'engineering' would allow the development of a practical silica-based TDFA. However, note that $Tm^{3+}$-doped silica fibre are very good for 2 µm laser sources (Jackson & King, 1999). This shows that the choice of optimal host glass is also dictated by the seeked application.

Other dopants have recently been proposed to explore amplification over new wavelength ranges. Bi-doped glasses with optical gain (Murata et al., 1999) and fibre lasers operating around 1100-1200 nm have been developed (Dianov et al., 2005; Razdobreev et al., 2007), although the identification of the emitting center is still not clear, and optimization of the efficiency is not yet achieved. Transition metal (TM) ions of the Ti-Cu series would also have interesting applications as broad band amplifiers, super-fluorescent or tunable laser sources, because they have in principle ten-fold spectrally larger and stronger emission cross-sections than RE ions. However, important NRD strongly reduces the emission quantum efficiency of TM ions in silica. Bi- and TM-doped fibres optical properties are extremely sensitive to the glass composition and/or structure to a very local scale. As we have shown, Cr ions in silica using standard fabrication methods provide strong and ultrabroad NIR luminescent, but only at low temperature (Dussardier et al., 2002). Interestingly, other application of TM-doped fibre have been proposed in lasers.

As for $Tm^{3+}$ ions, practical applications based on silica doped with alternative dopants along 'low efficiency' optical transitions would be possible when the 'ion site engineering' will be performed in a systematic approach. This approach is proposed via 'encapsulation' of dopants inside glassy or crystalline nanoparticles (NP) embedded in the fibre glass, like reported for oxyfluoride fibres (Samson et al., 2001) and multicomponent silicate fibres (Samson et al., 2002). In NP-doped-silica fibres, silica would act as support giving optical and mechanical properties to the fibre, whereas the dopant spectroscopic properties would be controlled by the NP nature. The NP density, mean diameter and diameter distribution must be optimized for transparency (Tick et al., 1995). In this context, our group has made contributions in various aspects introduced above. Our motivations are usually application oriented, but we address fundamental issues. First, the selected dopants act as probes of the local matrix environment, via their spectroscopic variations versus ligand field intensity, site structure, phonon energy, statistical proximity to other dopants, etc. The studies are always dedicated to problems or limitation in applications, such as for EDFA and TDFA, or high temperature sensors. It is also important to use a commercially derived fabrication technique, here the Modified Chemical Vapor Deposition (MCVD), to assess the potential of active fibre components for further development.

The aim of this chapter is reviewing of our contributions towards the comprehension and improvement of the spectroscopic properties of some RE and TM ions doped into silica. Its outline is as follow : Section 2 will describe the MCVD fabrication method of preform and fibre samples, and the characterization techniques applied to all samples. In Section 3, we report on the spectroscopic investigations of $Tm^{3+}$-doped fibres versus the material composition, including phonon interactions and non-radiative relaxations. It includes the proposal of potential applications by numerical simulations. Section 4 summarizes our original investigations on transition metals, focusing on chromium ions ($Cr^{3+}$ and $Cr^{4+}$) in silica-based fibres. Applications of $Cr^{4+}$-doped optical fibres as integrated saturable absorbers for passively Q-switched lasers are investigated. In section 5 are reported our recent discoveries in RE-doped dielectric nanoparticles, grown by phase separation. Finally, perspectives and conclusions are drawn in section 6.

## 2. Experimental

### 2.1 Preforms and fibres fabrication

All fibres investigated in this article were drawn from preforms prepared by the MCVD technique (Nagel et al., 1985) at the Laboratoire de Physique de la Matière Condensée (Nice). In this process, chemicals (such as $O_2$, $SiCl_4$) are mixed inside a silica tube that is rotating on a lathe. The flame from a burner translating along the lathe axis heats the tube and locally produces a chemical oxidizing reaction that transforms $SiCl_4$ and $O_2$ into $SiO_2$ molecules and gaseous chlorine. Extremely fine silica particles are deposited as soot on the inner side of the tube. This soot is transformed into a glass layer when the burner is passing over. The cladding layers are synthesized first, followed by the core layer(s). Germanium and phosphorus can also be incorporated directly through the MCVD process. They are added to raise the refractive index. Moreover, the latter serves also as a melting agent, decreasing the melting temperature of the glass. All the other elements like RE, TM, Al (a common glass modifier) are incorporated through the so-called solution doping technique (Townsend et al., 1987). The core layer is deposited at lower temperature than the preceding cladding layers, so that they are not fully sintered and left porous. Then the substrate tube is filled with an alcoholic ionic solution and allowed to impregnate the porous layers. After soaking the solution is removed, the porous layer is dried and sintered. The tube is then collapsed at 2000 °C into a cylindrical preform. The preform is drawn into a fibre using a vertical tube furnace on a drawing tower. The preform tip is heated above 2000°C and as the glass softens, a thin drop falls by gravity and pulls a thin glass fibre. The diameter of the fibre is adjusted by varying the pulling capstan speed, under controlled tension. The fibre is then coated with a UV-curable polymer and is finally taken up on a rotating drum.

### 2.2 Material characterizations

Refractive index profiles (RIP) of the preforms and fibres were measured using dedicated commercial refractive index profilers. The oxide core compositions of the samples were deduced from RIP measurements on preforms, knowing the correspondence between index rising and $AlO_{3/2}$, $GeO_2$, $PO_{5/2}$ concentration in silica glass from the literature. The composition was also directly measured on some preforms using electron probe microanalysis technique in order to compare results. The concentration of these elements is generally around few mol%. Luminescent ions concentrations are too low to be measured through the RIP. They were measured through absorption spectra. For example, the $Tm^{3+}$ ion concentration has been deduced from the 785 nm ($^3H_6$=>$^3H_4$) absorption peak measured in fibres and using absorption cross-section reported in the literature (Jackson & King, 1999) : $\sigma_{abs}$(785 nm) = 8.7x10$^{-25}$ m$^2$.

## 3. Thulium-doped fibres

### 3.1 Improvement of the 800 and 1470 nm $Tm^{3+}$ emission efficiencies

Thulium-doped fibres have been widely studied in the past few years. Because of $Tm^{3+}$ ion rich energy diagram (Fig. 1), lasing action and amplification at multiple infrared and visible wavelengths are allowed. In this paragraph, we will focus on the $^3H_4$ manifold. Thanks to the possible stimulated emission peaking at 1.47 µm ($^3H_4$ => $^3F_4$, see Fig. 1), discovered by (Antipenko, 1983), one of the most exciting possibilities of $Tm^{3+}$ ion is amplifying optical signal in the S-band (1.47–1.52 µm), in order to increase the available bandwidth for future optical communications. The $^3H_4$ level can also decay radiatively by emitting 800 nm radiation which is of primary importance for high power laser and medical applications. Unfortunately, the upper $^3H_4$ level of this transition is very close to the next lower $^3H_5$ level so NRD are likely to happen in high phonon energy glass hosts, causing detrimental gain quenching.Improvement of the Tm3+ spectroscopy was proposed through Tm-Tm (Simpson et al., 2006) and Yb-Tm (Simpson et al., 2008) energy transfers mechanims. In this paragraph, we discuss on the modification of the local phonon energy around $Tm^{3+}$ ions to reduce NRD. Then, we present the resulting improvement in emission efficiencies for specific applications through numerical modeling.

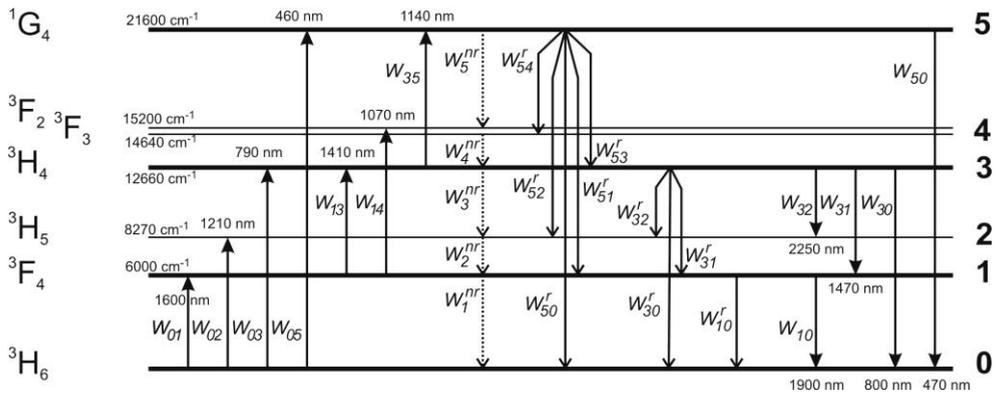

Fig. 1 : Energy diagram of $Tm^{3+}$ ion. Parameters are described in the text.

### 3.1.1 Local Phonon interactions

The relevant NRD transition of interest here is from the $^3H_4$ level downto the $^3H_5$ level. The NRD rate is expressed as (Van Dijk, 1983):

$$W_3^{nr} = W_0 \times \exp\left[-\alpha_{VD}\left(\Delta E - 2E_p\right)\right] \quad (1)$$

where $W_0$ and $\alpha_{VD}$ are constants depending on the material, $\Delta E$ is the energy difference between the both levels and $E_p$ is the phonon energy of the glass. To improve the quantum efficiency (which depends on the fluorescence lifetime) of the 800 and 1470 nm emission bands, modifications of $Tm^{3+}$ ion local environment were investigated by co-doping with selected modifying oxides (Faure et al., 2007). $GeO_2$ and $AlO_{3/2}$ have a lower maximum phonon energy than silica. As opposite demonstration, high phonon energy $PO_{5/2}$ was also used. To investigate the modification of the local environment, decay curves of the 810 nm fluorescence from the $^3H_4$ level were recorded. All the measured decay curves are non-exponential. This can be attributed to several phenomena and will be discussed in the next paragraph. Here, we study the variations of 1/e lifetimes ($\tau$) versus concentration of oxides of network modifiers (Al or P) and formers (Ge). The lifetime strongly changes with the composition of the glass host. The most striking results are observed within the *Tm(Al)* sample series: $\tau$ linearly increases with increasing $AlO_{3/2}$ content, from 14 µs in pure silica to 50 µs in sample *Tm(Al)* containing 17.4 mol% of $AlO_{3/2}$. The lifetime was increased about 3.6 times. The lifetime of the 20 mol% $GeO_2$ doped fibre *Tm(Ge)* was increased up to 28 µs whereas that of the 8 mol% $PO_{5/2}$ doped fibre *Tm(P)* was reduced down to 9 µs. Aluminum codoping seems the most interesting route among the three tested codopants.

### 3.1.2 Non-exponential shape of the 810-nm emission decay curves

In aluminium-doped fibres, fluorescence decay curves from the $^3H_4$ level were found to be non-exponential. It is thought that $Tm^{3+}$ ions are inserted in a glass which is characterized by a multitude of different sites available for the RE ion, leading to a multitude of decay constants. This phenomenological model was first proposed by Grinberg et al. and applied to $Cr^{3+}$ in glasses (Grinberg et al., 1998). This model was applied for the first time to $Tm^{3+}$-doped glass fibres (Blanc et al., 2008). In this method, the luminescence decay is given by:

$$I(t) \approx \sum_i A_i \exp\left[-t/\tau_i\right] \quad (2)$$

where $A_i$ and $\tau_i$ are discrete distributions of amplitudes and decay constants (lifetimes), respectively. For the fitting procedure on the $A_i$ series, 125 fixed values for $\tau_i$ were considered, logarithmically spaced from 1 to 1000 µs. For a given composition (Fig. 2a), we can notice two main distributions of the decay constant. With the aluminium concentration, they increase from 6 to 15 µs and from 20 to 100 µs, respectively (Fig. 2b). From the histograms of the amplitude distributions obtained from the fittings, characteristic lifetimes were correlated with those expected for thulium located either in a pure silica or pure $Al_2O_3$ environment. The $^3H_4$ lifetime is calculating by using this equation:

$$1/\tau = 1/\tau_{rad} + W_3^{nr} \tag{3}$$

where $\tau_{rad}$ corresponds to the radiative lifetime which is given to be 670 µs in silica (Walsh & Barnes, 2004). $W_0$ and $\alpha$ were estimated for different oxide glasses (Van Dijk & Schuurmans, 1983; Layne et al., 1977). The energy difference $\Delta E$ was estimated by measuring the absorption spectrum of the fibres. When Al concentration varies, this value is almost constant around 3700 cm$^{-1}$ (Faure et al., 2007). With these considerations, the $^3H_4$ expected lifetime can be calculated. In the case of silica glass, $\tau_{silica}$ = 6 µs and for a pure $Al_2O_3$ environment, $\tau_{alumina}$ = 110 µs. Both values agree with those obtained from the fitting procedure. The distribution of decay constant around 10 µs corresponds to $Tm^{3+}$ ions located in almost pure silica environment whereas the second distribution is attributed to $Tm^{3+}$ located in $Al_2O_3$-rich sites. This result shows that the global efficiency is increased by increasing the Al concentration in TDFAs. However concentrations above 20 at% would cause excess loss and glass stability problems in silica.

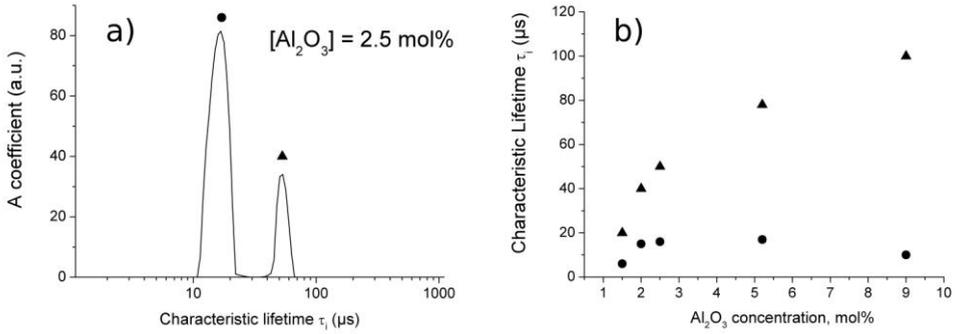

Fig. 2: a) Histogram of the recovered amplitude distributions obtained for silica-based $Tm^{3+}$-doped fibres. b) variations of the short (circle) and long (triangle) characteristic lifetimes vs $Al_2O_3$ concentration.

### 3.2 Modeling of Tm-doped silica-fibre devices

Numerical models are useful tools for investigating thulium-doped fibre devices, predicting their performance and optimization of their parameters. We have developed comprehensive, spectrally and spatially resolved numerical model that is based on simultaneous solution of the laser rate equations and propagation equations that describe evolution of the optical power along the fibre. The rate equations for the relevant energy levels can be written according to the energy level diagram in Fig. 1 as follows:

$$\frac{dn_1}{dt} = n_0(W_{01} + W_{02}) - n_1(W_{10} + W_{13} + W_{14} + W_1^{nr} + W_{10}^r) +$$
$$+ n_3(W_{31} + W_3^{nr} + W_{32}^r + W_{31}^r) + n_5(A_{51}^r + A_{52}^r), \tag{4}$$

$$\frac{dn_3}{dt} = n_0(W_{03} + W_{04}) + n_1(W_{13} + W_{14}) + n_5(W_5^{nr} + W_{54}^r + W_{53}^r) -$$

$$- n_3\left(W_{35} + W_{31} + W_{30} + W_3^{nr} + \sum_{j=0}^{2} W_{3j}^r\right), \quad (5)$$

$$\frac{dn_5}{dt} = n_0 W_{05} + n_3 W_{35} - n_5\left(W_{50} + W_5^{nr} + \sum_{j=0}^{4} W_{5j}^r\right), \quad (6)$$

while it holds that the sum of population $n_i$ on each respective level is equal to the total thulium ion concentration $n_t$ in the core. It is assumed that the thulium ions are homogenously distributed and excited in the doped section of radius $b$ in the fibre core of radius $a$ and it holds that $b \leq a$. The transition rates $W_{ij}$ accounts for the stimulated absorption and emission between the respective levels. Spontaneous decay processes are described by $W_{ij}^r$ and $W_i^{nr}$, the radiative and nonradiative decay rates, respectively. The formulae for the transition rates can be found in (Peterka et al., 2004). In steady state the rate equations become a set of four linear algebraic equations. The propagation of optical power at respective wavelength is governed by the following propagation equation:

$$\frac{dP^{\pm}(\lambda)}{dz} = \pm \Gamma(\lambda) P^{\pm}(\lambda) \sum_{ij}^{\{10,30,31,50\}} \left(n_i \sigma_{ij}(\lambda) - n_j \sigma_{ji}(\lambda)\right)$$

$$\Box \Gamma(\lambda) P^{\pm}(\lambda) \left(n_0 \sigma_{02}(\lambda) + n_0 \sigma_{04}(\lambda) + n_1 \sigma_{14}(\lambda) + n_3 \sigma_{35}(\lambda)\right)$$

$$\pm \Gamma(\lambda) \sum_{ij}^{\{10,30,31,50\}} 2h\nu_{ij} \Delta \nu n_i \sigma_{ij}(\lambda) \Box \alpha(\lambda) P^{\pm}(\lambda). \quad (7)$$

where $h$ is the Planck constant, $P^{\pm}$ are the spectral power densities of the radiation propagating in both directions along the fibre axis and $\sigma_{ij}$ is the respective transition cross section. The cross sections are shown in Fig. 3. The spectral dependence of the cross sections are approximated by a set of Gaussian functions, the respective coefficients can be found in (Peterka et al., 2011). The overlap factor $\Gamma$ represents the fraction of the transversal field distribution that interacts with the RE ions. The first term in equation (7) describes the amplification and reabsorption of optical signals, the second term represents ground-state absorption and excited state absorption (ESA) in spectral bands with no significant emission, the third term accounts for spontaneous emission and the fourth term stands for background loss $\alpha$ of the fibre. Evolution of optical power in each ASE spectral slot around a wavelength $\lambda$ (typical 1 nm wide slot is used) is governed by its respective propagation equation. The propagation equations together with the set of the rate equations under steady state conditions are solved simultaneously along the fibre using Runge-Kutta-Gill method of the fourth order. Since the boundary conditions for the counterpropagating partial waves $P^-$ are not known at the beginning of the fibre, an iterative solution is applied. The above described numerical model was verified by experimental comparisons (Blanc et al., 2006; Lüthi et al., 2007).

### 3.3 Applications

Thulium-doped fibres are renowned for their applications in high power fibre lasers at around 2 μm. It is despite the low quantum conversion efficiency of $^3F_4$ level in thulium-doped silica fibres, which is about 10% compared to ~100% quantum conversion efficiency of the ytterbium- and erbium-doped fibres, at around 1 μm and 1.5 μm, respectively. The lower quantum conversion efficiency increases the 2 μm laser threshold but has almost no effect on the laser slope efficiency. Indeed, kW-class thulium-doped fibre lasers have been demonstrated recently (Moulton, 2011). Quantum conversion efficiency of the $^3H_4$ level in non-modified silica fibres is much lower than that of $^3F_4$ level, only about 2% and therefore most of the applications of laser transitions originating from $^3H_4$ level are hindered by the lack of reliable low-phonon fibre host as discussed above. In the following two paragraphs we will show potential of the developed thulium-doped fibres with enhanced $^3H_4$ level lifetime for applications in fibre amplifiers for communication S-band and for fibre lasers around 810 nm.

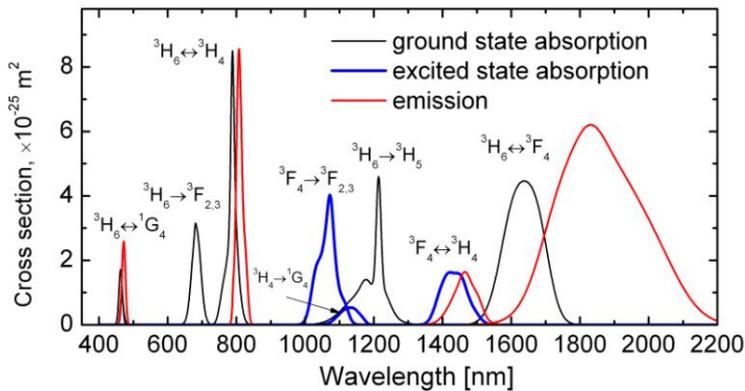

Fig. 3 : Absorption and emission cross section spectra of thulium.

### 3.3.1 Thulium-doped fibre amplifier at 1.47 μm

The low-loss and single-mode spectral range of the most common single-mode telecommunication fibre spans from about 1260 nm to 1675 nm. However, this tremendous transmission bandwidth is nowadays used for long-haul transmission only in a restricted portion in the C- (1530-1565 nm) and L-bands (1565-1625 nm), where reliable EDFA are available. The next logical frontier is the S-band (1460-1530 nm). One of the most promising candidates for amplification in the S-band is the TDFA. Together with EDFA it can substantially increase the bandwidths. An example of a TDFA structure is shown in Fig. 4.

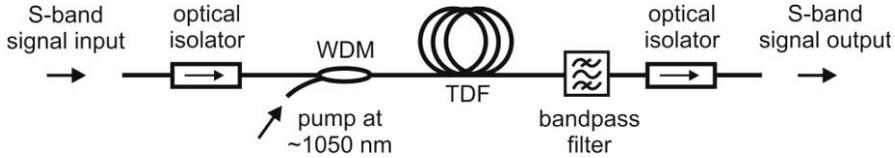

Fig. 4 : Typical layout of the TDFA.

Two main obstacles have to be solved in order to use the $^3H_4 \rightarrow {}^3F_4$ transition in the TDFA. Firstly, the quantum conversion efficiency of the $^3H_4$ thulium level should be increased, or in other words, the $^3H_4$ fluorescence lifetime should be made longer. One possible solution of the first problem is described above. Secondly, the lifetime of the upper laser level $^3H_4$ is shorter than that of the lower level $^3F_4$. Therefore, direct pumping of the $^3H_4$ level will hardly constitute the necessary population inversion. The second problem can be solved by using more complex pumping scheme employing gradual upconversion of the thulium ion to the upper laser level. It can be done by a single laser source as shown in Fig. 5a. The disadvantage is the possibility of the ESA $^3H_4 \rightarrow {}^1G_4$ that results in loosing pump photons. Many dual-wavelength pumping scheme have been also reported as reviewed in (Peterka et al., 2004). Several results of TDFA performance simulation and of optimization of the TDFA parameters are shown in Fig. 5 and Fig. 6. The TDFA spectral gain is shown in Fig. 5b for different silica fibre host represented by the fluorescence lifetime of the $^3H_4$ level: 14 μs of the typical non-modified silica fibre and 45 μs of the alumina-doped fibre developed by the authors and 55 μs of the highly germanium-oxide-doped fibre (Cole & Dennis, 2001). Similar value of 58 μs was also measured in highly alumina-doped fibre (Peterka et al., 2007). Codirectional pump of 1 W at 1064 nm is assumed. Unless otherwise stated, we consider the thulium concentration $n_t$=1.56×10$^{25}$ m$^{-3}$, the core diameter 2.6 μm, numerical aperture $NA$=0.3 and the radiative lifetimes of the levels $^3F_4$, $^3H_4$ and $^1G_4$ are 3500, 650 and 860 μs, respectively. The fluorescence lifetimes of the levels $^3F_4$, $^3H_4$ and $^1G_4$ are 430, 45 and 784 μs, respectively. For sake of comparisons we set the cross sections of the $^3F_4$ and $^1G_4$ levels the same for all fibre host types, although in real fibres they also depend on the host material. We have checked that their values have little effect on the calculated S-band gain compared to the effect of $^3H_4$ lifetime. The relevant branching ratios were estimated using Judd-Ofelt theory and were evaluated as follows: $\beta_{54}$=0.03, $\beta_{53}$=0.11, $\beta_{52}$=0.30, $\beta_{51}$=0.06, $\beta_{50}$=0.50, $\beta_{32}$=0.03, $\beta_{31}$=0.09, and $\beta_{30}$=0.88. Zero background loss is assumed. Signal power evolution along the thulium-doped fibre is shown in Fig. 5c for three input pump powers at 1064 nm. The available gain is significantly reduced by the presence of ASE, mainly around 800 nm. Suppression of the ASE, e.g., by cascaded inscription of long-period fibre gratings into the doped fibre or by using thulium-doped photonic crystal fibre with tailored band-gaps, would ameliorate the gain as shown also in Fig. 5.

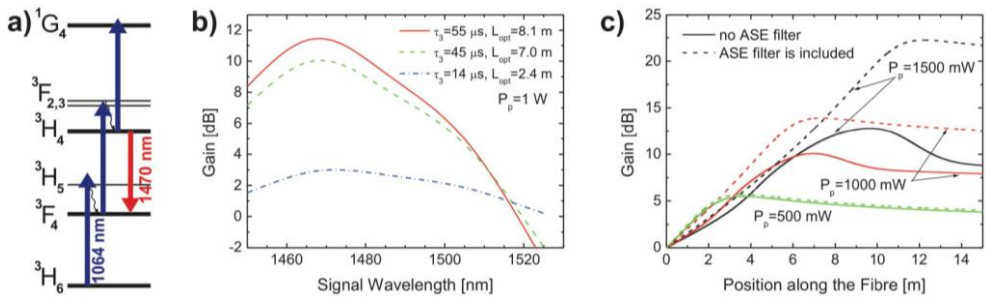

Fig. 5 : Upconversion pumping of TDFA at around 1050 nm (a). The effect of glass host composition, mainly manifested by the $^3H_4$ level lifetime, on the gain in the S-band (b) and optimization of the TDFA length (c).

Optimization of the fibre waveguide parameters is shown in Fig. 6a. Similarly as in the case of the EDFA, the gain gets higher with increasing numerical aperture and narrower core diameter. The effect of distributed filtering out of the ASE around 800 nm is also shown in the graph. Optimization of the pump wavelength is calculated in Fig. 6b. The optimum pump wavelength shifts towards shorter wavelength with increasing power levels. Shorter wavelength's pump sees lower GSA but at the same time higher first step ESA and consequently the integral value of the population inversion along the fibre is higher.

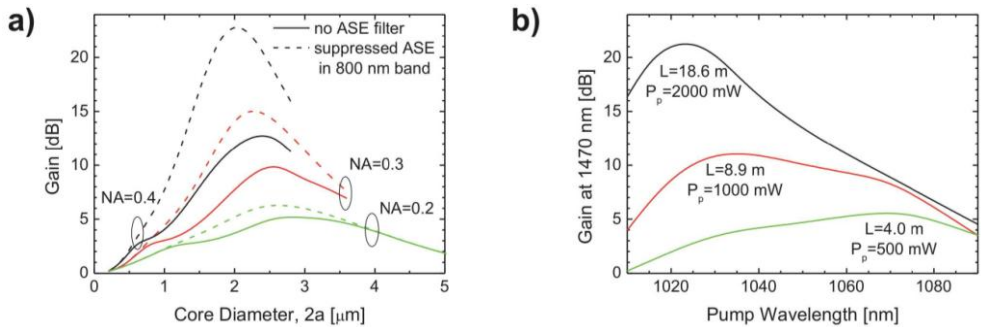

Fig. 6 : Optimization of the waveguide parameters, namely the NA and core radius (a) and optimiziation of the pump wavelength (b).

### 3.3.2  Thulium-doped fibre laser at 800 nm

The silica-based fibre lasers around 800 nm would extend the spectral range covered by high-power fibre lasers. The single-transversal mode, high-power laser source in the 800 nm spectral band is of interest for a variety of applications. The laser can be used for fibre sensors, instrument testing and for pumping of special types of lasers and amplifiers, e.g. the bismuth-doped lasers. Bismuth-doped fibres pumped around 800 nm may shift their gain to 1300 nm telecommunication band, where highly reliable silica-based fibre amplifiers are still unavailable. An efficient fibre laser in 800 nm spectral region could potentially be used as a replacement for titanium sapphire laser in some applications. A high-power amplifier in the 800 nm band would be useful in the optical fibre communications and short-haul free-space communications. Although the laser diodes at this wavelength have been available for a long time, to our knowledge, commercially available single-mode laser diodes are limited to about 200 mW of output power in diffraction limited beam. It should not be confused with, e.g., laser diode stacks of ~kW output power, that are highly multimode with very high $M^2$ >1000 factor. The amplification and lasing at 800 nm band has been already investigated using fluoride-based TDFs and output power of up to 2 W and 37% slope efficiency was achieved (Dennis et al., 1994). The output power was limited by the pump damage threshold of the fluoride fibre. Fibre-host reliability problems might be solved by using of silica-based fibres with enhanced $^3H_4$ lifetime. We proposed compact all-silica-fibre setup, example of which is shown in Fig. 7a. The proposed laser utilizes upconversion pumping scheme according to Fig. 7b.

We have performed optimization of the waveguide parameters, namely the core radius and NA (Peterka et al., 2011). The optimum length of the TDF can be determined from the calculated dependence of the laser output on the TDF length as it is shown in Fig. 7c for the three laser hosts. The laser output power vs. pump wavelength is shown in Fig. 7d. In comparison with the TDFA for the S-band telecommunications with optimal pump wavelength around 1020 nm (Fig. 6), the region of optimal pump wavelengths is shifted towards longer wavelength because in this case high inversion between $^3H_4$ and $^3H_6$ levels and high pump absorption from $^3H_6$ level is desirable. The background loss of 0.1 dB/m was considered. It can be seen in Fig. 7c that lasing at 810 nm is hard to achieve with silica-based Tm-doped fibre in contrast to the fluoride host materials. However, the lasing might be possible even for silica-based fibre for specific short range of the fibre lengths. Especially when the $^3H_4$ lifetime is enhanced, the laser output is very close to the one of ZBLAN host.

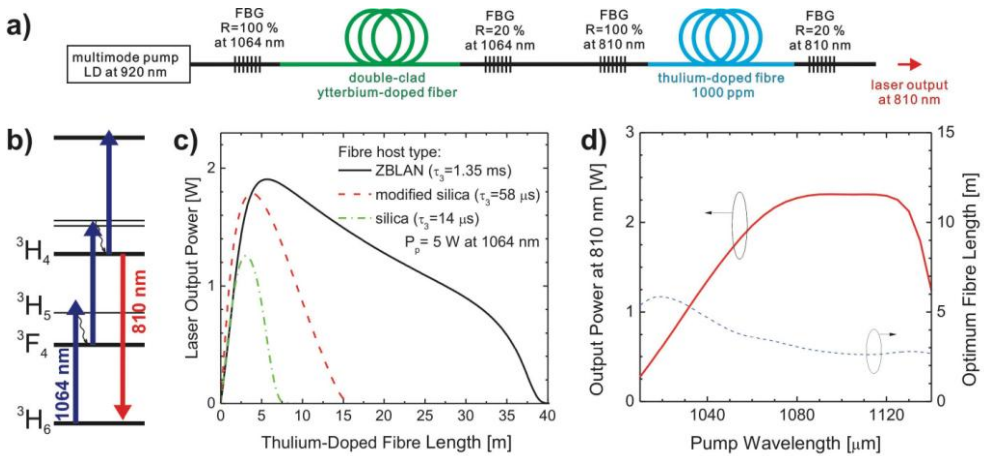

Fig. 7 Fibre laser setup in compact all-fibre arrangement (a), single wavelength up-conversion pumping scheme (b), effect of the host material (c) and pump wavelength (d). The waveguide parameters used for the parts (c) and (d) are: the core diameter of 3.4 μm and numerical aperture of 0.2.

## 3.4 Conclusions

Using a comprehensive numerical model we have shown the potential of the developed thulium-doped fibres for applications in fibre amplifiers for S-band telecommunication and for fibre lasers around 810 nm. The gain exceeding 20 dB of the S-band TDFA could be obtained with optimized fibre waveguiding parameters and optimized pump wavelength. Although the required pump power levels are relatively high, of the orders of Watts, thanks to the progress in ytterbium-doped fibre lasers even these pump powers can provide a cost effective solution. We have shown that efficient lasing at 810 nm can be achieved using silica-based Tm-doped fibre with enhanced $^3H_4$ lifetime for specific ranges of the fibre and laser cavity parameters. Such a fibre laser would broaden the spectral range currently covered by silica-based fibre lasers and it may replace the conventional laser systems in numerous applications.

## 4. Transition metal ions in silica-based optical fibres

### 4.1 Introduction

Many applications need ultra-broad band gain optical fibre materials. The well established tunability of RE-doped fibre devices is limited by shielding of the optically active electronic orbitals of RE ions, whereas unshielded orbitals are found in transition metal (TM) ions. One of the most promising is $Cr^{4+}$ because of its NIR ultra-broad band transitions. However incorporating of $Cr^{4+}$ ion in optical materials is difficult because it is less stable than the $Cr^{3+}$ ion. Some cristals such as $Cr^{4+}$:YAG are very good broad-band gain media (Sennaroglu et al., 1995), but little literature deals with $Cr^{4+}$-doped bulk glasses (Cerqua-Richardson et al., 1992; Hömmerich et al., 1994), and even fewer on Cr-doped silica fibres (Schultz, 1974; Chen et al.; 2007). Note that $Ni^{2+}$ in vitroceramic fibres is also a promising amplifying dopant (Samson et al., 2002). However, the proposed fibres lack of high transparency, reliability and cost-effectiveness of silica-based fibres, as produced by MCVD for instance. We have explored this field through basic studies on the optical properties of TM ions in silica optical fibres. In particular, the final TM oxidation states in the fibre core are strongly process-dependent. Also, the optical properties of one particular TM oxidation state (say $Cr^{4+}$) is difficult to interprete because they depend (1) on the host composition and final structure, due to crystal-field (or ligand field, in glass) fluctuations (Henderson & Imbush, 1989) and (2) on the experimental conditions (temperature, pressure, excitation wavelength,…). We describe the specific preparation details used for the Cr-doped fibres, we sum up spectroscopic results and interpretations, and finally we summarize our original studies on Cr-doped fibre saturable absorbers for all-fibre passively Q-switched (PQS) fibre lasers.

### 4.2 Fabrication and characterization of Chromium-doped samples

The fibres were prepared as in section 2.1, using $Cr^{3+}$ alcoholic doping solutions and oxidizing or neutral atmosphere for the drying-to-collapse stages. Samples containing Ge or/and Al were prepared, referred to as *Cr(Ge)*, *Cr(Ge-Al)* and *Cr(Al)*, respectively. We determined the absolute Cr content and relative concentrations of oxidation states using plasma emission and electron paramagnetic resonance spectroscopies, respectively. We assigned the optical transition of $Cr^{3+}$ and $Cr^{4+}$ using the Tanabe-Sugano (T.-S.) formalism (Sugano et al., 1970) and we qualitatively determined their strength, energy and bandwidth. From composition and sharing out of valencies, we have determined the absorption cross-sections. Absorption, emission and decay measurements were performed at room (RT) and low temperatures (LT, 12 or 77 K), using various pump wavelengths. Full details of the experimental procedures and analysis are given in (Felice et al., 2000, 2001; Dussardier et al., 2002).

### 4.3 Local structure, valency states and spectroscopy of transition metal ions

The main finding is that only $Cr^{3+}$ and $Cr^{4+}$ oxidation states were found. $Cr^{3+}$ is favoured by Ge co-doping in *O*-symmetry as in other oxide glasses (Henderson & Imbush, 1989 Rasheed et al., 1991), whereas $Cr^{4+}$ in distorted tetrahedral site symmetry ($C_s$) (Anino et al., 1997) is present in all samples, and is promoted by Al or when *[Cr]* is high (Fig. 8). The most interesting samples are the low-doped *Cr(Al)* ones containing only $Cr^{4+}$. The $Cr^{3+}$ absorption cross section $\sigma_{abs}^{3+}(670\ nm) = 43 \times 10^{-24}\ m^2$ is consistent with reports in other materials like ruby (Cronemeyer, 1966) and silica glass (Schultz, 1974), while $\sigma_{abs}^{4+}(1000\ nm) \sim 3.5 \times 10^{-24}\ m^2$ is lower than in reference crystals for lasers (Sennaroglu et al., 2006) or saturable absorbers (Lipavsky et al., 1999), but consistent with estimated values in alumino-silicate glass (Hömmerich et al., 1994).

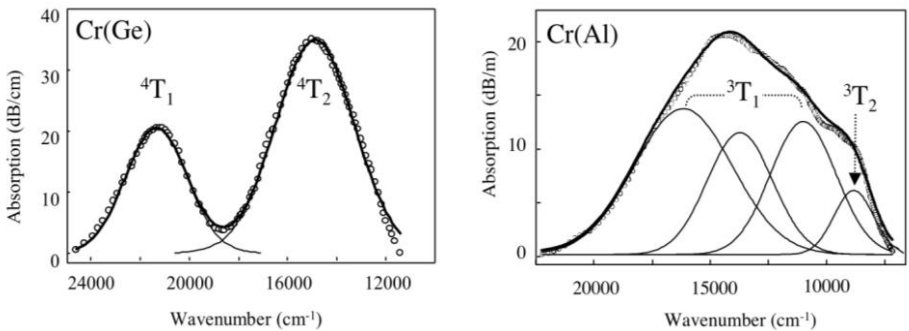

Fig. 8:. Absorption from a *Cr(Ge)* preform (*[Cr]* = 1400 mol ppm) and a *Cr(Al)* fibre (*[Cr]* = 40 mol ppm). Solid lines are gaussian fittings to $Cr^{3+}$ (left) and $Cr^{4+}$ (right) transitions, respectively. Assignments are from the ground level $Cr^{3+}:^4A_2$ or $Cr^{4+}:^3A_2$ to the indicated level, respectively. The $Cr^{4+}$ energy states are referred to by their irreductible representation in the $T_d$ symmetry coordination. The $Cr^{4+}:^3T_2$ level three-fold splitting is due to distorsion from perfect $T_d$ symmetry.

The main normalized parameter from the T.-S. formalism (*Dq/B=1.43*) was lower than the so-called 'crossing value' (*Dq/B* = 1.6), and lower than those reported for $Cr^{4+}$ in laser materials like YAG and forsterite (Anino et al., 1997). As expected consequences, a broad featureless NIR emission band along the $^3T_2 \rightarrow {}^3A_2$ transition is observed and no narrow emission line from the $^1E$ state is seen (Fig. 9, left). The LT fluorescence from $Cr^{4+}$ spreads from 850 to 1700 nm, and strongly varies depending on core chemical composition, *[Cr]* and $\lambda_p$ (pump wavelength). The observed bands were all attributed to $Cr^{4+}$ ions, in various sites. Fig. 9 shows examples of fluorescence spectra of $Cr^{4+}$ in different samples and experimental conditions. Possible emission from $Cr^{3+}$, $Cr^{5+}$ or $Cr^{6+}$ centers was rejected (Dussardier et al., 2002). The fluorescence sensitivity to *[Cr]* and $\lambda_p$ suggests that Cr-ions are located in various host sites, and that several sites are simultaneously selected by scanning of $\lambda_p$. It is also suggested that although Al promotes $Cr^{4+}$ over $Cr^{3+}$ when *[Cr]* is low, $Cr^{4+}$ is also promoted at high *[Cr]* in Ge-modified fibres. The strong decrease of fluorescence from LT to RT (not shown) is attributed to temperature quenching caused by NRD through multiphonon relaxations, like in crystalline materials where the emission drops by typically an order of magnitude from 77 K to 293 K (Sennaroglu et al., 1995).

The non-exponential LT fluorescence decays (Fig. 9, right) depend on *[Cr]* and $\lambda_s$. The fast decay part is assigned to Cr clusters or $Cr^{4+}$-rich phases within the glass. The 1/e-lifetimes ($\tau$) at $\lambda_s$ = 1100 nm are all within the 15-35 μs range in Al-containing samples, whereas $\tau \sim$ 3-11 μs in *Cr(Ge)* samples, depending on *[Cr]*. The lifetime of isolated ions ($\tau_{iso}$), measured on the exponential tail decay curves (not shown) reach high values: $\tau_{iso} \sim$ 200 to 300 μs at $\lambda_s \sim$ 1100 nm, $\tau_{iso} \sim$ 70 μs at $\lambda_s \sim$ 1400 nm. In heavily-doped *Cr(Ge)* samples, $\tau_{iso}$ is an order of magnitude less. Hence, $Cr^{4+}$ ions are hosted in various sites: the lowest energy ones suffer more NRD than the higher energy ones. Also presence of Al improves the lifetime, even at high *[Cr]*. It is estimated that at RT, lifetime $\tau$ would be much less than 1 μs.

As a summary of these spectroscopic studies, the main results are: (i) the observed LT fluorescence of $Cr^{4+}$ is extremely sensitive to glass composition, total *[Cr]* and excitation wavelength; (ii) using Al as a glass network modifier has advantages: longer excited state lifetime and broader fluorescence bandwidth than in Ge-modified silica; (3) the RT emission is strongly limited by NRD in silica. Broadband light sources would need a better control of the local environment around TM ions. However this study has allowed to propose an alternative application to Cr-doped fibre for self pulsing lasers.

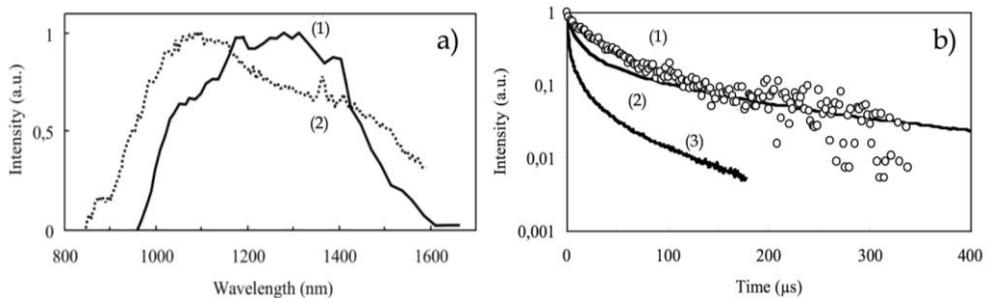

Fig. 9: Fluorescence spectra (a) : (1) fibre *Cr(Al)*, $\lambda_p$ = 900 nm, *T*=77 K ; (2) preform *Cr(Ge-Al)*, $\lambda_p$ = 673 nm, *T*=12K. Fluorescence decays (b) from *Cr(Al)* samples, $\lambda_p$ = 673 nm, *T*=12 K: (1) $\lambda_s \sim$ 1100 nm and *[Cr]*=40 ppm, (2) $\lambda_s \sim$ 1100 nm, *[Cr]*=4000 ppm, and (3) $\lambda_s \sim$ 1400 nm, *[Cr]*=4000 ppm.

## 4.4 Passively Q-switched fibre lasers using a $Cr^{4+}$–doped fibre saturable absorber.

The development of Q-switched fibre lasers is an alternative to bulk pulsed solid-state lasers operating in the ns- to µs-range for applications such as material processing and metrology. Actively Q-switched (AQS) fibre lasers have been extensively studied (Wang & Xu, 2007), as well as passively Q-switched (PQS) lasers (Siegman, 1986). Most AQS and PQS fibre lasers use externally driven bulk modulators or bulk passive components (Hideur et al., 2001; Paschotta et al., 1999) or intracavity TM-doped crystals (Laroche et al., 2002), located in a free space section of the cavity, causing alignment and reliability critical problems. To avoid them, we are interest into all-fibre systems comprising a saturable absorber (SA) fibre. We had experimentally demonstrated the first all-fibre PQS laser using a Cr-doped optical fibre saturable absorber (SA) (Tordella et al., 2003). Before, only one numerical study had mentioned this possibility (Luo & Chu, 1999), and 'auto' self-pulsing had been observed in heavily $Er^{3+}$-doped fibres (Sanchez et al., 1993). However, to obtain stable PQS behaviour, the criterium of 'second laser threshold' must be met (Siegman, 1986): It imposes that the SA absorption cross-section be larger than the gain medium emission cross section. Also, the SA absorption must spectrally overlap the emission gain curve and the SA relaxation time must be short to allow for high repetition rates. RE absorption transitions have limited spectral width and sometimes very long decay times. For these reasons, TM ions are in principle better SA elements: they offer broad absorption and fast decay. For example, $\sigma_{abs}^{4+}(1\ \mu m)$ is at least 10 times larger than $\sigma_{em}^{Nd3+}$ (Felice et al., 2000) and $\tau_{Cr4+}$ <1 µs is much less than of $\tau_{Nd3+}$(500 µs) or $\tau_{Yb3+}$ (800 µs) (Dussardier et al., 2002).

Our first all-fibre PQS laser, operating at 1084 nm, used a core-pumped Nd-doped fibre spliced to a Cr-doped fibre saturable absorber (CrSA) (Tordella et al., 2003). Although the pulse energy was low (~15 nJ), the characteristic behaviour agreed with theory, and the principle was demonstrated. Since, few PQS fibre lasers have implemented alternative SA ions in fibres, like RE ($Ho^{3+}$, $Tm^{3+}$, $Sm^{3+}$) or Bi, using sometimes rather complex cavity configurations, or heavily RE doped fibre SA (Kurkov, 2011). More recently, double-clad ytterbium-doped fibre (DCYF) lasers have raised a great interest. In some cases, DCYF lasers self-pulse in a random manner (Upadhyaya et al., 2010), along two main operating regimes: the sustained self-pulsing (SSP: period longer than the cavity round-trip time), and the self mode-locking (SML: period equal to the cavity round-trip time) (Brunet et al., 2005). The setup of our second system is schematically shown in Fig. 10. It is based on a home-made D-shaped DCYF. All experimental details are found in (Dussardier et al., 2011). The cavity parameters were chosen so that it would provide favorable conditions for SSP. As a consequence, the laser slope efficiency (SE) was not optimized. That also increased the threshold of appearance of stimulated Brillouin scattering (SBS) far above the achieved output power, therefore SBS had not effect on the reported results, and all the observed dynamic could be attributed to PQS mode of operation.

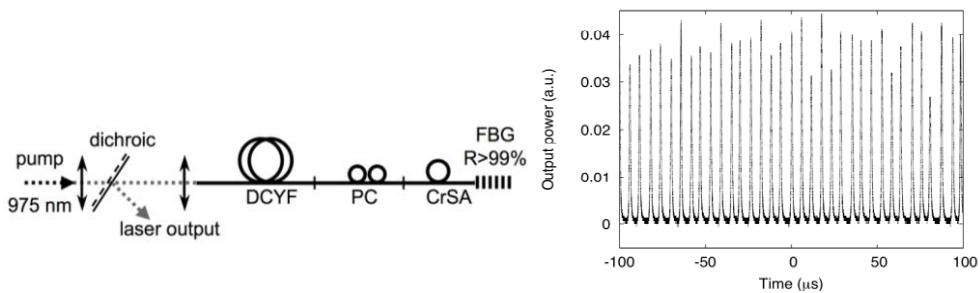

Fig. 10: Yb$^{3+}$:Cr$^{4+}$ PQS laser setup, and typical pulse train. DCYF: double-clad Yb-doped optical fibre, PC: polarization controler, CrSa: Cro-doped fibre saturable absorber, FBG: high reflectivity fibre Bragg grating mirror.

The DCYF laser without the CrSA had a chaotic behaviour, randomly switching between continuous-wave (CW) and Q-switched operation. The pulse envelopes (few µs long) were strongly modulated with a period of one cavity return-trip time (Fig. 11(a)), a sign of a combination of SSP and SML. The same was observed using a commercial Yb$^{3+}$-heavily doped fibre. The CrSA fibre drastically changed the laser characteristics. The µs-long pulse trains were stabilized over the whole available pump range (peak RMS fluctuations < 10 % at 9.4 times the threshold) (Fig. 10). The PQS mode was present at any pump power. The smooth Q-switched pulses did not show the sub-modulation at the return trip period that was visible in the 'DCYF only' laser (Fig. 11(b)). The reasons of the stabilization are not fully understood yet, and are still under study. The PQS laser threshold and SE were 1 W and 6 %, respectively. Although the SE has not been optimized it compares well with other fibre lasers using RE- or Bi-doped fibre SA (up to ~10%) (Kurkov, 2011). At 16 W of incident pump power the minimum pulse duration, highest repetition rate and peak power were 480 ns, 350 kHz and 20 W, respectively. Such a PQS laser source could be optimized to increase the intracavity optical density, and hence to saturate the CrSA more and faster, and produce more energetic pulses. It could also be a seed in a master-oscillator power amplifier configuration to be amplified by power Yb-doped fibre amplifiers.

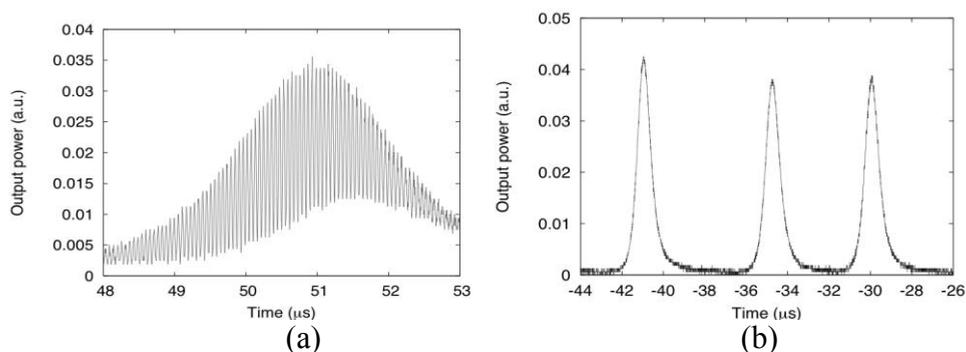

Fig. 11 : (a) zoom on a typical pulse envelope from the 'DCYF only' chaotic laser output. (b) zoom on few typical pulses from the stabilized 'DCYF+CrSA' laser.

### 4.5 Conclusion on the study of Cr-doped optical fibres.

Some glass modifiers like Al in silica-based fibres induce major spectroscopic changes, even at low concentrations (~1-2 mol%). This would help engineering the Cr optical properties in silica-based fibres, using possibly alternative modifiers. An interesting application for PQS all-fibre lasers was found. However, increasing of the quantum efficiency would be necessary for amplifiers and light sources. One needs to control the TM local environment by some material engineering. Preliminary work was reported (Dvoyrin et al., 2003) on post-heat treatment of $Cr^{3+}$-doped Ga-modified silica fibres. Other possible material engineering would be the nanoscale spontaneous phase separation.

### 5. Rare-earth-doped dielectric nanoparticles

Optical fibres are mostly based on silica glass for its interesting properties, as mentioned in section 1. However, some of its characteristics (high phonon energy, low solubility of luminescent ions, ...) induce drawbacks in spectroscopic properties of rare-earth (RE)-doped fibre lasers and amplifiers. Here we investigate the growing of RE-doped dielectric oxide nanoparticles (DNP) in silica-based optical fibres. Through this route, we keep the optical and mechanical advantages of silica glass, and we investigate the potential of engineering of RE spectroscopic properties through the choice of the DNP chemical composition.

Scarce reports on RE-doped transparent glass ceramics (TGC) singlemode fibres use low melting mixed oxides prepared by a rod-in-tube technique (Samson et al., 2002), or mixed oxyfluorides using a double-crucible technique (Samson et al., 2001), both with a subsequent ceramming stage. However the low melting point of these materials causes low compatibility with silica components. Transition metal-doped silica-based TGC fibres were prepared by MCVD (Modified Chemical Vapour Deposition) and using a slurry method (Yoo et al., 2003), i.e. the particles were synthesized before insertion into the silica tube-substrate.

In this section, we discuss a more straightforward and original technique to embed RE ions within *in-situ* grown oxide nanoparticles in silica-based preforms (Blanc et al., 2009a). The implemented principle is the spontaneous phase separation process (Zarzycki, 1991). Silicate systems can exhibit strong and stable immiscibility when they contain divalent metals oxides (MO, where M= Mg, Ca or Sr) (Fig. 12). For example, if a silicate glass containing few mol% MO is heated it will decompose into two phases : one silica-rich and one MO-rich in shape of spherical particles. Two key advantages of this process are that (i) nanoparticles are grown *in-situ* during the course of the fabrication process and (ii) there is no need (and associated risks) of nanoparticles manipulation by an operator. Further, the process takes advantage of the high compositional control and purity typical of the MCVD technique. The so-called 'solution doping technique' was applied to incorporate alkaline-earth and erbium ions. To raise the core refractive index and ease the fabrication, germanium (~2 mol%) and small amounts of phosphorus (~1 mol%) were added. In the next paragraphs, we discuss the influence of the alkaline-earth ions on the nanoparticles formation. We also present the attenuation measurements as well as the nanoparticles composition. Finally, modifications of the erbium spectroscopy are described.

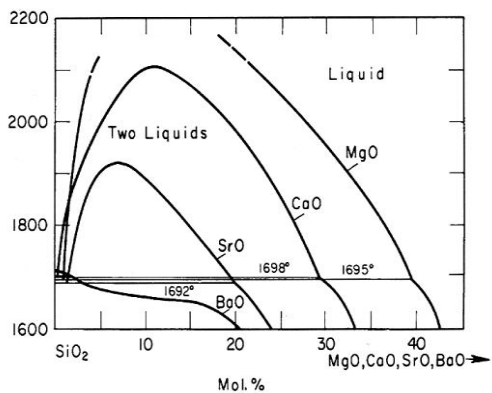

Fig. 12: Immiscibility-gap in the phase-diagram of binary $SiO_2$-MO glass (M = Mg, Ca, Sr)

### 5.1  Effect of Mg, Ca and Sr on the formation of the nanoparticles

When alkaline-earth (Mg, Ca and Sr) ions are incorporated, nanoparticles are observed both in preforms and fibres. Typical SEM pictures from the exposed core section of cleaved fibres are shown on Fig. 13 (Mg, Ca and Sr concentrations in the doping solution are 0.1 mol/l). The gray disk corresponds to the fibre core (~8 µm). The dark central part of the core is caused by the evaporation of germanium element; this is a common artifact of the MCVD technique that can be corrected through process optimization. Nanoparticles are visible as bright spots. They show an important compositional contrast compared to the silica background. The size distribution of the NP depends strongly on the nature of the alkaline earth ions. It is about 50 nm for Mg-doped fibre while it reaches 100 nm or more for Sr- and Ca-doped samples. The size of the nanoparticles is also dependent on the concentration of the alkaline-earth ions (Blanc et al., 2011). For the fibre doped with the solution with 0.1 mol/l of $MgCl_2$, the mean particle size is 48 nm and no particle bigger than 100 nm was observed in these MgO-doped fibres, unlike the case of CaO-doped fibres. Analysis of the SEM images also revealed that the inter-particle distance is in the 100 - 500 nm range. When the solution concentration increases from 0.1 to 1 mol/l of $MgCl_2$, the inter-particle distance remains nearly the same but the mean particle diameter almost doubles to reach 76 nm. With this concentration, NPs up to 160 nm were observed.

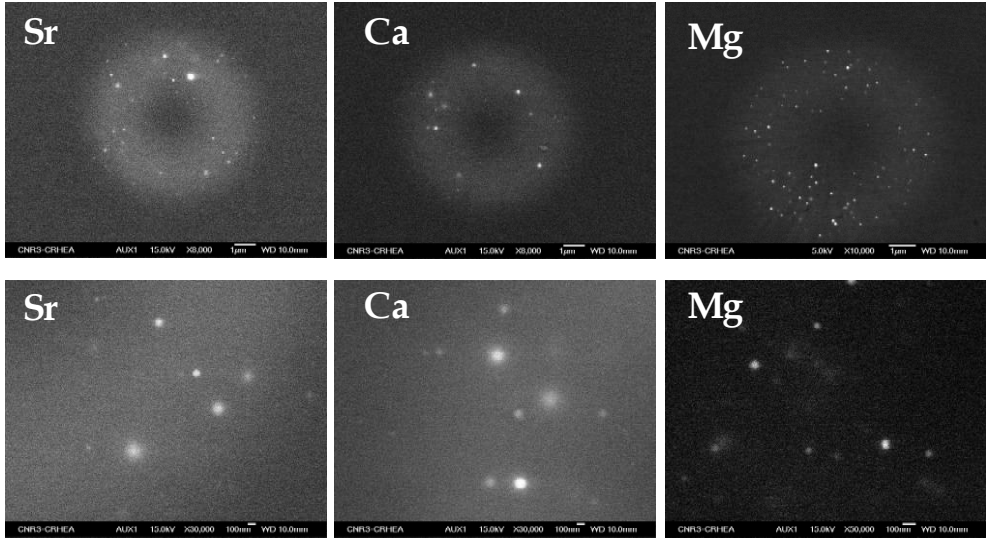

Fig. 13: SEM pictures of a MO-doped fibre (M = Sr, Ca, Mg). For each fibres, alkaline-earth concentration is 0,1 mol/l.

### 5.2 Transparency

The scattering loss in Mg-doped fibre was measured through the cut-back method. The fibre bending radius was kept > 20 cm to minimize bend loss. The 0.1 mol/l Mg-doped fibre attenuation spectrum is displayed in Fig. 14. At the wavelength of 1350 nm, losses were measured to be 0.4 dB/m only. This value is comparable with the attenuation measured in low-melting temperature transparent glass ceramics fibres (Samson et al., 2002) and is compatible with amplifier applications. The loss variation vs wavelength is attributable to light scattering induced by the nanoparticles. Due to the small particle size we assume Rayleigh scattering to be the major source of scatter loss, which was estimated according to the formula: $\alpha$ (dB/m)=4.34.$C_{Rayleigh}$.$N$.$\Gamma$, where $N$ is the DNP density (m$^{-3}$), $\Gamma$ is the overlap factor between the field and the core containing the DNP ($\Gamma$ = 0.3 in fibre A) and $C_{Rayleigh}$ (m²) is the Rayleigh scattering coefficient (Bohren & Huffman, 2004):

$$C_{Rayleigh} = \frac{(2\pi)^5}{48} \times \frac{d^6}{\lambda^4} \times n_m^4 \times \left( \frac{n_n^2 - n_m^2}{n_n^2 + 2n_m^2} \right)^2 \quad (8)$$

where $d$ is the nanoparticle diameter, $n_m$ and $n_n$ the host material and particles refractive indices, respectively. The actual particle composition is not known but a high content of MgO is expected (Stebbins et al., 2009). The nanoparticle refractive index is estimated at ~ 1.65, like that of Mg-based oxide such as $Mg_2SiO_4$ (Burshtein et al., 2003). Under these considerations, fitting of the experimental data with Eq. 8 yields a particle density $N \sim$ 0.4 x 10$^{20}$ m$^{-3}$, or equivalently a mean inter-particle distance ~300 nm. These values agree well with the results collected from the SEM pictures taken from cleaved fibres.

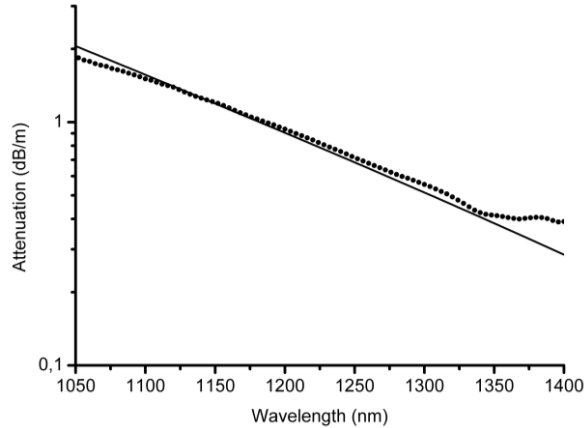

Fig. 14: Transmission spectrum of a Mg-doped fibre (0.1 mol/l solution). Dots: experimental data, full line: calculated Rayleigh scattering curve.

At 1350 nm, the normalized frequency V was lower than 1.3 ($LP_{11}$ mode cut-off wavelength is 700 nm). The difference between experimental data and Rayleigh scattering curve above 1350 nm is attributed to bending loss only. However, for practical application, the presented fabrication technique allows any necessary waveguide optimization without preventing the TGC growth. The attenuation of other fibres (1 mol/l of $MgCl_2$ or Ca- and Sr-doped) was extremely high (several 100s dB/m), in agreement with the Rayleigh formula (120 dB/m) assuming large monodisperse 100 nm particles. This shows that the particle mean size must be less than 50 nm for potential applications, and that the technique presented here is able to produce fibres with acceptable scattering loss.

### 5.3 Composition of the nanoparticles

The composition was investigated by EDX analyses. When nanoparticles are analyzed, Ca or Mg, P and Si are found while only Si is detected outside of the particles. Germanium seems to be homogeneously distributed over the entire glass. When erbium is added to the composition, it is found to be inserted into the particles as it is presented on Fig. 15. Such conclusion was also drawn when emission intensity of $Er^{3+}$ was mapped on the end-face of a cleaved fibre under a confocal microscope (Blanc, 2009b). No erbium fluorescence is detected outside the DNP. This result was expected due to the low solubility of RE ions in silica (or even in germano-silicate), whereas strongly modified and amorphous silicates, such as in the DNP, have a high solubility for these ions. These results indicate that erbium ions are located inside or very close to the DNP.

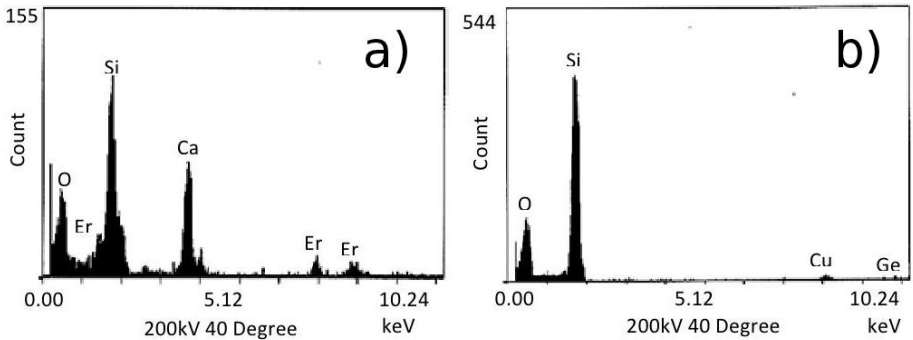

Fig. 15: Energy dispersive x-ray analyses spectra of the preform sample doped with Ca and Er. The area analysed corresponds to the nanoparticle (a) and outside (b)

### 5.4 Erbium spectroscopy

The emission spectra and lifetime from fibres doped with 0.1 (Fibre A) and 1 mol/l (Fibre B) were measured at room temperature around 1.55 µm under 980-nm pump excitation. The emission spectrum from Fibre A (Fig. 16) is similar to that in a silicate environment. Fibre B shows a distinct broadening of its spectrum (FWHM is 44 nm) by as much as ~ 50 % compared to Fibre A. In comparison, more than 10 at% of aluminum, as network modifier, would have been necessary to obtain the same FWHM in 'type III' Er-doped fibres in optical amplifiers for telecommunications (Desurvire, 1994), because Al and Er are evenly distributed across the core volume. Moreover, the shape of the fluorescence spectrum from Fibre B is quite unusual: it decreases monotonically between the peaks at 1.53 and 1.55 µm and the commonly observed dip at 1.54 µm is absent. These features would be attractive for realizing intrinsically gain flattened fibre amplifiers, provided sufficient minimization of the scattering loss is ensured through process optimization. Both fibres A and B produced single-exponential decaying fluorescence with 11.7 and 6.7 ms lifetimes, respectively.

Modifications of the $Er^{3+}$ spectroscopic properties in the TGC optical fibres are clearly evident from the above results when the Mg concentration increases. Such observations were previously reported in Ca-doped preforms. The broadening of the fluorescence curve and the lower lifetime obtained from Fibre B is tentatively interpreted as an effect of the modification of the erbium ions averaged local field. In other words, erbium ions in Fibre B are, on average, located in a medium with stronger local field compared to Fibre A. This induces stronger spontaneous transition probability, and hence a shorter lifetime (Henderson, 1989). Although the exact composition of the NPs is not yet known, emission spectrum and fluorescence lifetime of erbium ions in Fibre B are closely related to the results reported in the literature in phosphate glasses (Liu et al., 2003 ; Jiang et al., 1998). When Mg concentration increases, erbium ions environment changes from silicate to phosphate.

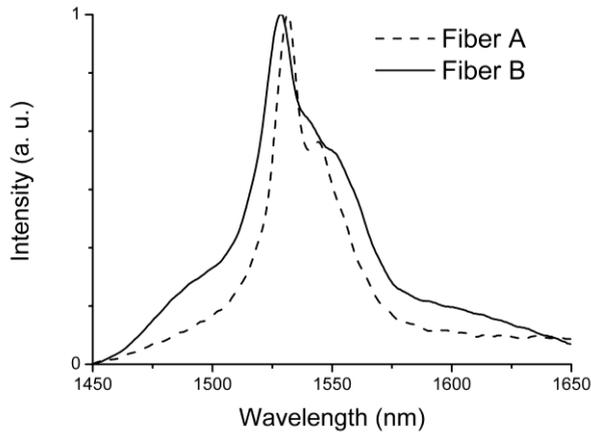

Fig. 16: Room temperature emission spectra from Mg-doped fibres. The Mg concentration in the doping solution is 0.1 (A) and 1.0 (B) mol/l, respectively. Excitation wavelength is 980 nm.

## 5.5   Conclusion

A method to fabricate $Er^{3+}$-doped TGC fibers entirely through the MCVD process is demonstrated. By adding magnesium to the silica-based composition, a low-loss fiber is obtained through *in situ* growth without requiring a separate process to realize NP (such as postprocess ceramming). An important result of this study is that the type of obtained nanoparticle by this technique could be as large as ~50 nm for applications such as in fiber amplifiers and lasers. Moreover, a broadening of the emission spectrum by as much as ~50% is observed with attractive features to realize gain flattened fiber amplifiers. More generally, this concept might have great potential as a possible solution to address various current issues in amplifying fibers, including (but not limited to) realization of intrinsically gain flattened amplifiers, etc.

## 6.   General conclusion

The choice of a glass to develop new optical fibre component is most of the time a result of compromises. Silica glass is the most widely used for its many advantages (reliability, low cost fabrication, …). The overall composition suffers from different drawbacks, such as high phonon energy or low luminescent ions solubility, which affect quantum efficiency or emission bandwidth of luminescent ions, for example. However, spectroscopic properties of active ions are not directly related to the average properties, but to their local environment. Then, new luminescent properties of dopants can be obtained by slightly modifying the silica composition.

In this chapter, we have shown that thulium emission efficiency can be improved by reducing the local phonon energy. Using a comprehensive numerical model we have shown that efficient S-band TDFA and lasing at 810 nm can be obtained. The chromium valence state was controlled by adding some glass modifiers like Al. This open the way to PQS all-fibre lasers. Then, nanostructuration of doped fibre is proposed as a new promising route to 'engineer' the local dopant environment. All these results will benefit to optical fibre components such as lasers, amplifiers and sensors, which can now be realized with silica glass.

## 7.     References

Anino, C. ; Théry J. & Vivien, D. (1997). New $Cr^{4+}$ activated compounds in tetrahedral sites for tunable laser applications. *Optical Materials*, Vol. 8, No. 1-2, pp 121-128.

Antipenko, B.M. ; Mak, A.A. ; Raba, O.B. ; Seiranyan, K.B. & Uvarova, T.V. (1983). New lasing transition in the $Tm^{3+}$ ion, *Soviet Journal of Quantum Electronics*, Vol. 13, No 4, pp. 558-560.

Blanc, W. ; Peterka, P. ; Faure, B. ; Dussardier, B. ; Monnom, G. ; Kasik, I. ; Kanka, J. ; Simpson, D. & Baxter, G. (2006). Characterization of thulium-doped silica-based optical fibre for S-band amplifier. *SPIE 6180 Proceedings of Photonics, Devices, and Systems III*. ISBN: 80-86742-08-3, Prague (Czech Republic), June 2005.

Blanc, W. ; Sebastian, T.L. ; Dussardier, B. ; Michel, C. ; Faure, B. ; Ude, M. & Monnom, G. (2008). Thulium environment in a silica doped optical fibre. *Journal of non-crystalline solids.* Vol. 354, No 2-9, pp. 435-439.

Blanc, W. ; B. Dusssardier & M.C. Paul (2009a). Er-doped oxide nanoparticles in silica-based optical fibres, *Glass Technology: European Journal of Glass Science and Technology.* Vol. 50, No 1, pp. 79-81.

Blanc, W. ; Dussardier, B. ; Monnom, G. ; Peretti, R. ; Jurdyc, A.-M. ; Jacquier, B. ; Foret, M. & Roberts, A. (2009b). Erbium emission properties in nanostructured fibers. *Applied Optics* Vol. 48, No 31, pp. G119-124.

Blanc, W. ; Mauroy, V. ; Nguyen, L. ; Bhaktha, S.N.B. ; Sebbah, P. ; Pal, B.P. & Dussardier, B. (2011). Fabrication of rare-earth doped transparent glass ceramic optical fibers by modified chemical vapor deposition, *Journal of the American Ceramic Society.* DOI: 10.1111/j.1551-2916.2011.04672.x, published on-line.

Bohren, C.F. & Huffman, D.R..(2004). *Absorption and scattering of light by small particles,* Wiley Science, ISBN 0471293407, New-York.

Brunet, F. ; Taillon, Y. ; Galarneau, P. & LaRochelle, S. (2005). A simple model describing both self-mode locking and sustained self-pulsing in ytterbium-doped ring fiber lasers. IEEE *Journal of Lightwave Technology*, Vol.23, pp. 2131-2138.

Burshtein, Z. & Shimony, Y. (2002). Refractive index dispersion and anisotropy in $Cr^{4+}:Mg_2SiO_4$, *Optical materials,* Vol. 20, No. 2, pp 87-96.

Cerqua-Richardson, K. ; Peng, B. & Izumitani, T. (1992). Spectroscopic Investigation of $Cr^{4+}$-doped glasses. *OSA Proceedings of Advanced Solid-State Lasers*, ISBN 1557522235, Santa Fe (NM, USA, February 1992), Vol. 13. pp 52-55.

Chen, J.-C. ; Lin, Y.-S. ; Tsai, C.-N. ; Huang, K.-Y. ; Lai, C.-C & Su, W.-Z. (2007). 400-nm-bandwidth emission from a Cr-doped glass fiber. *IEEE Photonics Technology Letters*, Vol. 19, No. 8, pp 595 - 597.

Cole, B. & Dennis, M. L. (2001). S-band amplification in a thulium doped silicate fiber. *Proceedings of Conference on Optical Fiber Communication*, ISBN 1557526559, Anaheim (CA, USA), March 2001.

Cronemeyer, D.C. (1966). Optical absorption characteristics of pink ruby. *Journal of the Optical Society of America*, Vol. 56, No. 12, pp 1703-1705.

Dennis, M. L.; Dixon, J. W. & Aggarwal, I. (1994). High power upconversion lasing at 810 nm in Tm:ZBLAN fibre, *Electronics Letters*, Vol. 30, No. 2, pp. 136-13

Desurvire, E. (1994). *Erbium Doped Fiber Amplifiers: principles and applications*, Wiley Interscience, ISBN 0-471-58977-2, New-York.

Desurvire, E. ; Bayart, D. ; Desthieux, B. & Bigo, S. (2002). *Erbium Doped Fiber Amplifiers: device and system developments*, Wiley Interscience, ISBN 0-471-41903-6, New-York.

Dianov, E.M. ; Dvoyrin, V.V. ; Mashinsky, V.M. ; Umnikov, A.A. ; Yashkov, M.V. & Gur'yanov, A.N. (2005). CW bismuth fiber laser. *Quantum Electronics.* Vol. 35, No. 12, pp 1083-1084.

Digonnet, M. (2001). *Rare-Earth-Doped Fiber Lasers and Amplifiers* (2nd ed.) Marcel Dekker. ISBN 0-203-90465-6, New-York.

Durteste, Y. ; Monerie, M. ; Allain, J.-Y. & Poignant, H. (1991). Amplification and lasing at 1.3 µm in praseodymium-doped fluoridezirconate fibers. *Electronics Letters*, Vol. 27, pp. 626-628.

Dussardier, B. ; Guyot, Y. ; Felice, V. ; Monnom, G. & Boulon, G. (2002). $Cr^{4+}$-doped silica-based optical fibers fluorescence from 0.8 µm to 1.7 µm. *Proceedings of Advanced Solid State Lasers*, in Trends in Optics and Photonics Series (OSA), ISBN: 1-55752-697-4, Vol. 68, pp. 104-108.

Dussardier, B; Maria, J. & Peterka, P. (2011). Passively Q-switched ytterbium and chromium doped fibre laser. *Applied Optics*, Vol. 50, No. 25, pp. E20–E23.

Dvoyrin, V.V. ; Mashinsky, V. ; Neustruev, B. ; Dianov, E.M. ; Guryanov, A.N. & Umnikov, A.A. (2003). Effective room-temperature luminescence in annealed chromium-doped silicate optical fibers. *Journal of the Optical Society of America B*, Vol. 20, No. 2, pp. 280-283.

Faure, B. ; Blanc, W. ; Dussardier, B. & Monnom, G. (2007). Improvement of the $Tm^{3+}$:$^3H_4$ level lifetime in silica optical fibers by lowering the local phonon energy. *Journal of Non-Crystalline Solids.* Vol. 353, No 29, pp. 2767-2773.

Felice, V.; Dussardier, B. ; Jones, J.K. ; Monnom, G. & Ostrowsky, D.B. (2000). $Cr^{4+}$-doped silica optical fibres : absorption and fluorescence properties. *The European Physical Journal-Applied Physics*, Vol. 11, pp. 107-110.

Felice, V.; Dussardier, B. ; Jones, J.K. ; Monnom, G. & Ostrowsky, D.B. (2001). Chromium-doped silica optical fibres : influence of the core composition on the Cr oxidation states and crystal field. *Optical Materials*. Vol. 16, No. 1-2, pp. 269-277

Grattan, K.T.V. & Sun, T. (2000). Fiber optic sensor technology: an overview. *Sensors and Actuators A*: Physical, Vol. 82, No. 1-3, pp. 40-61.

Grinberg, M. ; Russell, D.L. ; Holliday, K. ; Wisniewski, K. & Koepke, Cz. (1998). Continuous function decay analysis of a multisite impurity activated solid. *Optics Communications*, Vol. 156, pp. 409–418.

Henderson, B. and Imbush, G.F. (1989). *Optical Spectroscopy of Inorganic Solids*, Clarendon Press, ISBN 0-19-851372-0, Oxford.

Hewak, D.W. ; Deol, R.S. ; Wang, J. ; Wylangowski, G. ; Mederios Neto, J.A. ; Samson, B.N. ; Laming, R.I. ; Brocklesby, W.S. ; Payne, D.N. ; Jha, A. ; Poulain, M. ; Otero, S. ; Surinach, S. & Baro., M.D. (1993). Low phonon-energy glasses for efficient 1.3 µm optical fiber amplifiers. *Electronics Letters*, Vol. 29, No. 2, pp. 237-239.

Hideur, A. ; Chartier, T. ; Brunel, M. ; Salhi, M. ; Özkul, C. & Sanchez, F. (2001). Mode-lock, Q-switch and CW operation of an Yb-doped double-clad fiber ring laser. *Optics Communications*, Vol. 198, No. 1-3, pp. 141–146.

Hömmerich, U. ; Eilers, H. ; Yen, W.M. ; Hayden, J.S. & Aston, M.K. (1994). Near infrared emission at 1.35 µm in Cr doped glass. *Journal of Luminescence*, Vol. 60&61, pp. 119-122.

Jackson, S.D. & King, T.A. (1999). Theoretical modelling of Tm-doped silica fiber lasers. *Journal of Lightwave Technology*. Vol. 17, No. 5, pp. 948-956.

Jiang, S. ; Myers, M. & Peyghambarian, N. (1998). $Er^{3+}$ Doped Phosphate Glasses and Lasers. *Journal of Non-Crystalline Solids.* Vol. 239, No 1-3, pp. 143-148.

Komukai, T. ; Yamamoto, T. ; Sugawa, T. & Miyajima, Y. (1995). Upconversion pumped Thulium-doped fluoride fiber amplifier and laser operating at 1.47 µm. *IEEE Journal of Quantum Electronics*. Vol. 31, No. 11, pp. 1880-1889.

Kurkov, A. S. (2011). Q-switched all-fiber lasers with saturable absorbers. *Laser Physics Letters*, Vol. 8, No. 5, pp. 335-342 + references therein.

Laroche, M. ; Chardon, A.M. ; Nilsson, J. ; Shepherd, D.P. ; Clarkson, W.A. ; Girard, S. & Moncorgé, R. (2002). Compact diode-pumped passively Q-switched tunable Er-Yb double-clad fiber laser. *Optics Letters*, Vol. 27, No. 22, pp. 1980-1982.

Layne, C.B. ; Lowdermilk W.H. & Weber, M.J. (1977). Multiphonon relaxation of rare-earth ions in oxide glasses. *Physical Review B*. Vol. 16, No 1, pp.10-21.

Lin, H. ; Tanabe, S. ; Lin, L. ; Hou, Y.Y. ; Liu, K. ; Yang, D.L. ; Ma, T.C. ; Yu, J.Y. & Pun, E.Y.B. (2007). Near-infrared emissions with widely different widths in two kinds of $Er^{3+}$-doped oxide glasses with high refractive indices and low phonon energies. *Journal of Luminescence*, Vol. 124, No. 1, pp. 167-172.

Lipavsky, B. ; Kalisky, Y. ; Burshtein, Z. ; Shimony, Y. & Rotman, S. (1999). Some optical properties of $Cr^{4+}$-doped crystals. *Optical Materials*, Vol. 13, No. 1, pp. 117-127.

Liu, Z. ; Qi, C. ; Dai, S. ; Jiang, Y. & Huet, L. (2003). Spectra and Laser Properties of $Er^{3+}$, $Yb^{3+}$:Phosphate Glasses. *Optical Materials*. Vol. 21, No 4, pp. 789-794.

Luo, L. & Chu, P.L. (1999) Passive Q-switched erbium-doped fibre laser with saturable absorber, *Optics Communications*, Vol. 161, No. 4-6, pp. 257 – 263.

Lüthi, S. R.; Gomes, A. S. L.; Sundheimer, M. L.; Dussardier, B; Blanc, W. & Peterka, P. (2007) Distributed gain in a Tm-doped silica fiber: experiment and modeling. Proceedings of CLEO-Europe, ISBN 1424409306, Munchen (Germany), June 2007.

Minelly, J. & Ellison, A. (2002). Applications of antimony-silicate glasses for fiber optic amplifiers. *Optical Fiber Technology*, Vol. 8, No. 2, pp. 123-138.

Moulton, P. M. (2011). High power Tm:silica fiber lasers: current status, prospects and challenges, Proceedings of CLEO-Europe, ISBN : 978-1-4577-0532-8, Munchen (Germany), May 2011

Murata, K. ; Fujimoto, Y. ; Kanabe, T. ; Fujita, H. & Nakatsuka, M. (1999). Bi-doped $SiO2$ as a new laser material for an intense laser. *Fusion Engineering and Design*. Vol. 44, No. 1-4, pp. 437-439.

Nagel, S.R. ; MacChesney, J.B. & Walker, K.L. (1985). *Modified chemical vapor deposition. in : Optical Fiber Communications: Vol 1 'Fiber Fabrication'*, ed. T. Li, pp 1-64, Academic Press, ISBN 0-12-447301-6, Orlando.

Paschotta, R. ; Häring, R. ; Gini, E. ; Melchior, H. ; Keller, U. ; Offerhaus, H. L. & Richardson, D. J. (1999). Passively Q-switched 0.1-mJ fiber laser system at 1.53 µm. *Optics Letters*, Vol. 24, pp. 388–390.

Peterka, P. ; Faure, B.; Blanc W.; Karasek, M. & Dussardier, B. (2004). Theoretical modelling of S-band thulium-doped silica fibre amplifiers. *Optical and Quantum Electronics*, Vol. 36, No. 1-3, pp. 201-212.

Peterka, P. ; Kasik, I.; Dhar, A.; Dussardier, B. & Blanc, W. (2011). Theoretical modeling of fiber laser at 810 nm based on thulium-doped silica fibers with enhanced $^3H_4$ level lifetime. *Optics Express*, Vol. 19, No. 3, pp. 2773-2781.

Peterka, P. ; Kasik, I.; Matejec V.; Blanc W.; Faure, B.; Dussardier, B.; Monnom, G. & Kubecek, V. (2007). Thulium-doped silica-based optical fibers for cladding-pumped fiber amplifiers, *Optical Materials*, Vol. 30, No. 1, pp. 174-176.

Peterka, P.; Blanc, W.; Dussardier, B.; Monnom, G.; Simpson, D. A. & Baxter, G. W. (2008). Estimation of energy transfer parameters in thulium- and ytterbium-doped silica fibers. SPIE 7138 Proceedings of Photonics, Devices, and Systems IV, ISBN: 978-081947379-0, Prague (Czech Republic), august 2008.

Rasheed, F. ; O'Donnell, K.P. ; Henderson, B. & Hollis, D.B. (1991). Disorder and the optical spectroscopy of $Cr^{3+}$-doped glasses: I. Silicate glasses. *Journal of Physics: Condensed Materials.* Vol. 3, No. 12, pp. 1915-1930.

Razdobreev, I. ; Bigot, L. ; Pureur, V. ; Favre, A. ; Bouwmans, G. & Douay, M. (2007). Efficient all-fiber bismuth-doped laser. *Applied Physics Letters.* Vol. 90, pp. 031103 (3 pages).

Samson, B.N. ; Pinckney, L.R. ; Wang, J. ; Beall, G.H. & Borrelli, N.F. (2002). Nickel-doped nanocrystalline glass-ceramic fiber. *Optics Letters.* Vol. 27, No. 15, pp 1309-1311.

Samson, B.N. ; Tick, P.A. & Borrelli, N.F. (2001). Efficient neodymium-doped glass-ceramic fiber laser and amplifier. *Optics Letters.* Vol. 26, No. 3, pp. 145-147.

Sanchez, F. ; Le Boudec, P. ; François, P.-L. ; Stephan, G. (1993). Effects of ion pairs on the dynamics of erbium-doped fiber lasers. *Physical Review A.* Vol. 48, No. 3, pp. 2220-2229.

Schultz, P.C. (1974). Optical Absorption of the Transition Elements in Vitreous Silica. *Journal of the American Ceramics Society.* Vol. 57, No. 7 pp. 309-313.

Sennaroglu, A. ; Demirbas, U. ; Ozharar, S. & Yaman, F. (2006). Accurate determination of saturation parameters for $Cr^{4+}$-doped solid-state saturable absorbers. *Journal of the Optical Society of America B*. Vol. 23, No. 2, pp. 241-249.

Sennaroglu, A. ; Pollock, C.R. & Nathel, H. (1995). Efficient continuous-wave chromium-doped YAG laser. *Journal of the Optical Society of America B.* Vol. 12, No. 5, pp. 930-937.

Siegman, A.E. (1986) *Lasers*, Univ. Science Books, ISBN 0-935702-11-5, Mill Valley.

Simpson, D.A. ; Baxter, G.W. ; Collins, S.F. ; Gibbs, W.E.K. ; Blanc, W. ; Dussardier, B. & Monnom, G. (2006). Energy transfer up-conversion in $Tm^{3+}$-doped silica fiber. *Journal of Non-Crystalline Solids.* Vol. 352, No 2, pp. 136-141.

Simpson, D.A. ; Gibbs, W.E.K. ; Collins, S. F. ; Blanc, W. ; Dussardier, B. ; Monnom, G. ; Peterka, P. & Baxter, G. W. (2008). Visible and near infra-red up-conversion in Tm$^{3+}$/Yb$^{3+}$ co-doped silica fibers under 980 nm excitation. *Optics Express.* Vol. 16, No 18, pp. 13781-13799.

Stebbins, J.F. ; Kim, N. ; Andrejcak, M.J. ; Boymel, P.M. & Zoitos B.K. (2009). Characterization of phase separation and thermal history effects in magnesium silicate glass fibers by nuclear magnetic resonance spectroscopy. *Journal of the American Ceramic Society.* Vol. 92, No 1, pp. 68-74.

Sugano, S. ; Tanabe, Y. & Kamimura, H. (1970) *Multiplets of Transition-Metal Ions in Crystals*, Academic Press, New York.

Tick, P.A. ; Borrelli, N.F. ; Cornelius, L.K. & Newhouse, M.A. (1995). Transparent glass ceramics for 1300 nm amplifier applications. *Journal of Applied Physics.* Vol. 78, No. ll, pp 6367-6374.

Townsend, J.E. ; Poole, S.B. & Payne, D.N. (1987). Solution doping technique for fabrication of rare earth doped optical fibers. *Electronics Letters.* Vol. 23, No. 7, pp. 329-331.

Upadhyaya, B. N. ; Chakravarty, U. ; Kuruvilla, A. ; Oak, S.M. ; Shenoy, M.R. & Thyagarajan, K. (2010). Self-pulsing characteristics of a high-power single transverse mode Yb-doped CW fiber laser. *Optics Communications.* Vol. 283, No. 10, pp. 2206-2213.

Van Dijk, J.M.F. & Schuurmans, M.F.H. (1983). On the nonradiative and radiative decay rates and a modified exponential energy gap law for 4f–4f transitions in rare-earth ions. *The journal of Chemical Physics.* Vol. 78, No 9, pp. 5317-5313.

Walsh, B.M. & Barnes, N.P. (2004). Comparison of Tm:ZBLAN and Tm:silica fiber lasers: Spectroscopy and tunable pulsed laser operation around 1.9 µm. *Applied Physics B.* Vol. 78, No 3-4, pp. 325-333.

Wang, Y. & Xu, Ch.-Q. (2007). Actively Q-switched fiber lasers: Switching dynamics and nonlinear processes (A review). *Progress in Quantum Electronics.* Vol. 31, No. 3-5, pp. 131–216.

Yoo, S. ; Paek, U.-C. & Han, W.-T. (2003). Development of a glass optical fiber containing ZnO–Al$_2$O$_3$–SiO$_2$ glass-ceramics doped with Co$^{2+}$ and its optical absorption characteristics. *Journal of Non-Crystalline Solids.* Vol. 315, No 1-2, pp. 180–186.

Zarzycki, J (1991). *Glasses and the vitreous state*. Cambridge university press, ISBN 0521355826, Cambridge.